# Electron-Hole Entanglement in a Quantum Spin Hall Insulator

Koji Sato, Mircea Trif, and Yaroslav Tserkovnyak

*Department of Physics and Astronomy, University of California, Los Angeles, California 90095, USA*

We demonstrate that entangled electron-hole pairs can be produced and detected in a quantum spin Hall insulator with a constriction that allows for a weak inter-edge tunneling. A violation of a Bell inequality, which can be constructed in terms of low-frequency nonlocal current-current correlations, serves as a detection of the entanglement. We show that the maximum violation of a Bell inequality can be naturally achieved in this setup, without a need to fine tune tunneling parameters. This may provide a viable route to producing spin entanglement in the absence of any correlations and pairing, where spin-to-charge conversion is enabled by the helical edge structure of a quantum spin Hall insulator.

The quantum theory of many-body systems introduces the concept of entanglement, which is manifested through nonlocal correlations that cannot be explained by any local hidden-variables theory. This unique feature of quantum mechanics allows for various new applications, such as quantum computation [1], quantum teleportation [2], and quantum cryptography [3–5]. As a consequence of the nonlocal correlations engendered by the entanglement, Bell inequalities can be constructed [6, 7], whose quantum-mechanical violation has been observed unambiguously in optical experiments [8–12]. On the other hand, the (nonoptical) production, manipulation, and detection of entangled particles in solid-state systems turned out to be challenging experimentally, due to their interaction with the surroundings, which affect the entanglement through decoherence. At the same time, ironically, these interactions can also serve as a resource for producing entanglement.

There are a number of schemes proposed for creating entangled states in solid-state systems, the vast majority focusing on entangling the spin degrees of freedom, as they are usually more robust against decoherence than their orbital counterparts. A natural way to generate spin-entangled electrons is by utilizing the interaction between electrons, such as the Coulomb interaction [13–15] in quantum dots or in the Fermi sea [16], or the pairing interaction in superconductors. As for the manipulation of entanglement, Cooper-pair (CP) splitters have been proposed, where spin-singlet CP's can be spatially separated by coupling an *s*-wave superconductor to different solid-state systems [17–20]. Moreover, orbital entanglement can also be produced utilizing superconductors [21] and quantum Hall bars [22, 23]. The detection of entanglement in these solid-state systems is based on the violation of a Bell inequality constructed with certain combinations of nonlocal current-current correlations (i.e., noise) [24, 25], in lieau of the coincidence measurements performed in optical experiments.

Remarkably, the set-ups in Refs. [22, 23] provide means to produce entangled electron-hole pairs via tunneling in the absence of any interactions, as opposed to all other examples. In this paper, we show that a quantum spin Hall insulator (QSHI) bar can supersede quantum Hall systems to both efficiently produce and detect entangled electron-hole pairs in the absence of any interactions and, importantly, without fine tuning of the tunneling characteristics. A schematic of our proposal is depicted in Fig. 1.

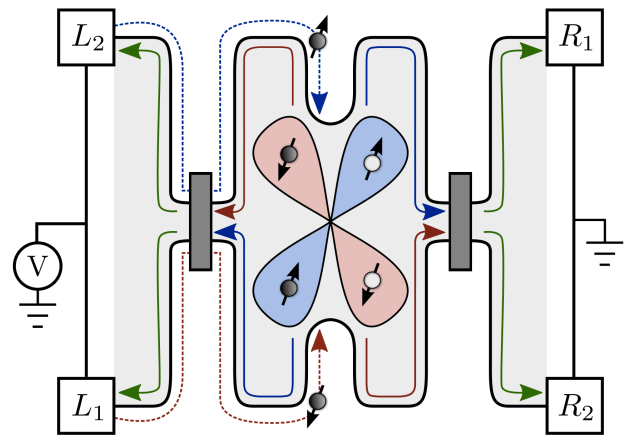

FIG. 1. An entangled electron-hole pair can be injected across the QSHI edges through the constriction in the middle, by biasing the left reservoir relative to the right reservoir. Blue (red) lobe indicates an electron-hole pair created by spin-up (down) incoming state from the left reservoir, denoted by blue (red) dashed trajectories. The entangled electrons and holes then propagate along the edges toward the two beam splitters (grey regions). The edge currents are finally probed at Fermi-liquid reservoirs $L_{1,2}$ and $R_{1,2}$.

Two-dimensional topological insulators [26], or QSHI's, have been recently realized experimentally in the inverted-band HgTe quantum-well heterostructures [27]. The edge states of a QSHI are robust against time-reversal invariant perturbations, and their spins and momenta are locked to each other in a helical fashion. A given edge of a QSHI hosts a pair of Kramers conjugates counter-propagating gapless modes with opposite spins, known as helical edge states. Due to this feature of the helical edge states, a QSHI could provide robust means for spin-to-charge conversion and thus electrical detection of spin entanglement. For instance, a CP splitter utilizing such helical edge states as charge carriers was proposed [28], where it was shown that the entangled spin-singlet state of CP's is reflected in current-current correlations. Such a CP splitter utilizing helical modes was recently proposed to perform a Bell test based on the current correlations along the edges of two QSHI's [29].

We consider a QSHI strip with a constriction formed in the middle, as shown in Fig. 1, where tunneling between the top and bottom edges can occur. Since the electrons at the edges are helical, the top (bottom) edge has spin-up (down) right-moving and spin-down (up) left-moving states. We denote by $a_{u,\uparrow(\downarrow)}$ Fermi-level fields for the right-(left-)moving incoming scattering state on the upper edge and by $a_{l,\downarrow(\uparrow)}$ the right-(left-)moving incoming state on the lower edge. Similarly, the outgoing states are given by fields $b_{u,\uparrow(\downarrow)}$ on the upper edge and $b_{l,\downarrow(\uparrow)}$ on the lower edge. The incoming and outgoing scattering states to the left (L) and right (R) of the scattering region can be related by a $4 \times 4$ unitary scattering matrix according to

$$\begin{pmatrix} \boldsymbol{b}_L \\ \boldsymbol{b}_R \end{pmatrix} = S \begin{pmatrix} \boldsymbol{a}_L \\ \boldsymbol{a}_R \end{pmatrix}, \quad (1)$$

where $\boldsymbol{a}_L = \left(a_{u,\uparrow} \ a_{l,\downarrow}\right)^T$, $\boldsymbol{a}_R = \left(a_{l,\uparrow} \ a_{u,\downarrow}\right)^T$, $\boldsymbol{b}_L = \left(b_{l,\uparrow} \ b_{u,\downarrow}\right)^T$, and $\boldsymbol{b}_R = \left(b_{u,\uparrow} \ b_{l,\downarrow}\right)^T$. In the following, we assume that the system is invariant under time reversal (no external magnetic field, magnetic impurities, etc.), which dictates: $T\boldsymbol{a} = ST\boldsymbol{b}$, where $\boldsymbol{a} = (\boldsymbol{a}_L, \boldsymbol{a}_R)^T$ and $\boldsymbol{b} = (\boldsymbol{b}_L, \boldsymbol{b}_R)^T$ are the incoming and outgoing states appearing in Eq. (1) and the time-reversal operator is given by $T = i\sigma_y K$, where $\sigma_y$ is the $y$ component of Pauli matrices (acting in the spin $\uparrow, \downarrow$ space of the scattering states within the $L, R$ blocks) and $K$ is the complex conjugation operator. By combining both the unitarity and the time-reversal symmetry of the scattering matrix, we thus obtain the relation $S^T = \sigma_y S \sigma_y$. This condition restricts the general expression for the scattering matrix to the following form [30] (see Supplementary Material (SM) [31]):

$$S = \begin{pmatrix} \hat{r} & -\hat{t}^\dagger \\ \hat{t} & \hat{r}^\dagger \end{pmatrix}, \quad (2)$$

where $\hat{r} = re^{i\phi_3}I$ and $\hat{t} = \tau Q$, with $I$ being the $2 \times 2$ unit matrix (we disregarded a global phase factor). Here, $r$ and $\tau$ are real-valued and satisfy $r^2 + \tau^2 = 1$, and the matrix $Q$ is given by

$$Q = \begin{pmatrix} e^{i\phi_1}\cos\theta & e^{i\phi_2}\sin\theta \\ e^{-i\phi_2}\sin\theta & -e^{-i\phi_1}\cos\theta \end{pmatrix}. \quad (3)$$

Here, $\theta$, $\phi_i$, with $i = 1, 2, 3$ are all real numbers parameterizing the scatterer, while the unitarity of $S$ implies

$QQ^\dagger = 1$. For an inversion symmetric constriction, the scattering matrix satisfies also $S^\dagger = \tau_x S \tau_x$, with $\tau_x$ being the $x$ component of the Pauli matrix acting in the $L, R$ blocks. This relation pertains to the extra condition $\hat{t} = \hat{t}^\dagger$ on the scattering matrix [31]. We will treat electrons as noninteracting, returning only in the end to comment on the possible sources of decoherence, including Luttinger-liquid correlations, that may reduce or destroy the entanglement.

We suppose the 2D system is biased such that the right reservoir is filled up to the Fermi energy $E_F$, while the left up to $E_F + eV$, with $V$ being the applied voltage bias. We are focusing on the low-temperature limit, $k_B T \ll eV$, and thus neglect thermal noise, in the following. The nonequilibrium current is composed of the right-moving states incident from the left reservoir above $E_F$. The only relevant incoming states are, therefore, $\boldsymbol{a}_L$, with the corresponding outgoing scattered states given by $\boldsymbol{b}_L = \hat{r}\boldsymbol{a}_L$ and $\boldsymbol{b}_R = \hat{t}\boldsymbol{a}_L$. Furthermore, the left-(right-)moving scattered states $\boldsymbol{b}_L$ ($\boldsymbol{b}_R$) impinge on the beam splitters located to the left (right) of the constriction. The beam splitters perform a unitary rotation by mixing the upper and lower edge states: $\boldsymbol{b}_{L(R)} \to U_{L(R)}\boldsymbol{b}_{L(R)}$, where the rotation matrix is given by

$$U_{L(R)} = \begin{pmatrix} \cos\frac{\theta_{L(R)}}{2} & \sin\frac{\theta_{L(R)}}{2} \\ -\sin\frac{\theta_{L(R)}}{2} & \cos\frac{\theta_{L(R)}}{2} \end{pmatrix}. \quad (4)$$

Here, the angles $\theta_{L(R)}$ parametrize scattering in the right (left) beam splitters, which can be controlled by gating, for example. The beam splitters are assumed to be much longer than the wavelength of the injected electrons, so that we can neglect the backscattering that may take place over their lengths.

In the following, we consider the electron-hole entanglement emergent in the present setup reflected through a viable Bell test (or Bell inequality). As shown in Ref. [22], such a test can be implemented in terms of the zero frequency current-current correlations of the helical edge states [24, 25] (defined below). However, we leave the details for such a connection to the SM [31], as the derivation follows closely the line of arguments in Ref. [22]. The frequency-dependent current correlation is found by Fourier transforming $S_{ij}(t) = \langle \{\delta I_{Li}(t), \delta I_{Rj}(0)\} \rangle / 2$, where $I_{L(R)i}$ is the current entering lead $L_i$ ($R_i$), with $i = 1, 2$ (see Fig. 1). Here, $S_{ij}(t)$ is the current-current correlator and $\delta I_{L(R)i} = I_{L(R)i} - \langle I_{L(R)i} \rangle$ is the fluctuation of the current. In the DC limit, the low-frequency noise is given by [34]

$$S_{ij} = -(e^3 V/h) \left| \left(\hat{r}\hat{t}^\dagger\right)_{ij} \right|^2 \equiv -(e^3 V/h)(r\tau)^2 |Q_{ij}^\dagger|^2, \quad (5)$$

which can be used to implement a Bell test [32]. To this end, it is convenient to define a quantity $E$ in terms of the following combination of the zero-frequency noise powers

[22]:

$$E \equiv \frac{S_{11} + S_{22} - S_{12} - S_{21}}{S_{11} + S_{22} + S_{12} + S_{21}} = \frac{\mathrm{Tr}(\sigma_z \hat{r}^\dagger \hat{t}^\dagger \sigma_z \hat{r} \hat{t}^\dagger)}{\mathrm{Tr}(\hat{r}^\dagger \hat{r} \hat{t}^\dagger \hat{t})}. \quad (6)$$

Next we take into account the effect of the beam splitters on the function $E$. The left-moving outgoing states through the beam splitter can be obtained by rotating the reflection matrix by $\hat{r} \to U_L^\dagger \hat{r} U_L$. Similarly, the right-moving outgoing states can be found by rotating the transmission matrix $\hat{t} \to U_R \hat{t} U_L$. In order to construct a Bell inequality, each beam splitter needs to be independently switched between two distinct rotation angles $\theta_{L(R)}$ and $\theta'_{L(R)}$, with the corresponding rotation matrices denoted by $U_{L(R)}$ and $U'_{L(R)}$. A Bell inequality (known as the Clauser-Horne-Shimony-Holt inequality [7]) $|B| \leq 2$, where

$$B \equiv |E(\theta_L, \theta_R) + E(\theta'_L, \theta_R) + E(\theta_L, \theta'_R) - E(\theta'_L, \theta'_R)|, \quad (7)$$

would indicate destruction of the entanglement by some external degrees of freedom that are outside of our treatment [33]. The maximum value of $B$ is given by $B_{\max} = 2\sqrt{1 + C^2}$ [22], with $C \in [0, 1]$ being the concurrence (see SM [31]). Hence, any nonzero value of the concurrence $C$ implies a violation of the Bell inequality fomented by electron-hole entanglement. While the general expression for $B$ is long and unilluminating, in the weak tunneling regime ($|r| \ll 1$ and $\theta \approx 0$) and in the presence of inversion symmetry, we obtain $E(\theta_L, \theta_R) = \cos(\theta_R - \theta_L)$, so that $B$ becomes

$$B = |\cos(\theta_L - \theta_R) + \cos(\theta'_L - \theta_R) + \cos(\theta_L - \theta'_R) - \cos(\theta'_L - \theta'_R)|, \quad (8)$$

reaching its maximum $B_{\max} = 2\sqrt{2}$ for, say, $\theta_L = 0$, $\theta_R = \pi/4$, $\theta'_L = \pi/2$, and $\theta'_R = -\pi/4$. The production of electron-hole entanglement based on a quantum Hall bar in Ref. [22] relies on having two integer quantum Hall channels. In the tunneling regime, the concurrence is given by $C = 2\sqrt{\tau_1 \tau_2}/(\tau_1 + \tau_2)$, where $\tau_{1,2}$ are the eigenvalues of the transmission $\hat{t}^\dagger \hat{t}$ corresponding to the two channels at the tunneling site, thus the maximum value of $C = 1$ is achieved when $\tau_1 = \tau_2$. On the other hand, these two channels are not identical and $\tau_1$ and $\tau_2$ are exponentially sensitive to the distances between the relevant edge states [35]. This is not an issue for the quantum spin Hall insulator, because $\tau_1 = \tau_2 = \tau$ is always satisfied according to Eq. (2), as a result of the time-reversal symmetry. Thus, the present setup naturally leads to $C = 1$ in the absence of any external tuning, which is one of the main results of this paper.

We note in passing that particle-hole entanglement can be produced even with a single Landau level, in a Hanbury Brown-Twiss proposal of Ref. [23]. Their setup, however, is more complex, requiring extra gates and leads, which may make the entanglement more susceptible to the decoherence sources in the external circuitry of the device. With the advent of single-particle emitters [36], our setup could also be extended to generate entanglement through a deterministic gate-controlled pumping [37] of discrete electron-hole pairs in the center of our schematic in Fig. 1, in the absence of a DC bias. A similar, albeit more complex, implementation was proposed in Ref. [38] for a QSHI-based pumping of electron pairs, which also bears some resemblance to Cooper-pair injection of Ref. [29].

Next we touch upon the loss of entanglement due to different sources of decoherence. At finite temperatures, the entanglement is affected by thermal fluctuations, which, however, can be shown to be irrelevant as long as $k_B T \ll eV$ [35]. There are two other important sources of decoherence that may result in the loss of entanglement: random phase shifts in the scattering matrix and intrinsic orbital dephasing due to electron-electron interactions within the helical edge channels. It was shown previously that entanglement is robust against random-phase fluctuations [35], as long as their amplitude stays below a critical value, while interaction effects can be accounted for within Luttinger-liquid phenomenology [39]. In particular, the charge fractionalization furnished by Luttinger-liquid correlations causes dephasing at finite temperature when $T > T_0 \equiv \hbar v / 2\pi k_B L$, $v$ being plasmon velocity and $L$ distance from the tunneling region in the center to beam splitters. In this high-temperature regime, the effective concurrence $\widetilde{C}$ suffers exponential decay as $\widetilde{C} \propto e^{-2\gamma T/T_0}$ [here, $\gamma \equiv (g + g^{-1})/2 - 1$ is the single-particle tunneling density-of-states exponent in a bulk Luttinger liquid, in terms of the interaction parameter $g$]. In the low-temperature limit, $T < T_0$, $\widetilde{C}$ follows power-law scaling characteristic of Luttinger liquids [39]. It was found that the total zero-frequency noise in the presence of interactions reads

$$\widetilde{S}_{ij} = e\bar{I}\left[1 + (-1)^{i+j}\left(\cos\theta_L \cos\theta_R + \widetilde{C}\sin\theta_L \sin\theta_R\right)\right]. \quad (9)$$

where $4\bar{I}$ is the average current flowing between the reservoirs ($\bar{I} \propto V$). Substituting this expression in Eq. (6) it results in the following expression for $E$:

$$E(\theta_L, \theta_R) = \cos\theta_L \cos\theta_R + \widetilde{C}\sin\theta_R \sin\theta_L. \quad (10)$$

The Bell inequality remains the same as in Eq. (7), but now accounting for the reduced concurrence $\widetilde{C}$. The noninteracting ($g = 1$) zero-temperature case gives the maximally-entangled result with $\widetilde{C} = 1$ with $B_{\max} = 2\sqrt{2}$.

Even in the presence of dephasing, i.e., $\widetilde{C} < 1$, by adjusting the four angles, $\theta_L$, $\theta'_L$, $\theta_R$, and $\theta'_R$, the maximum value of the Bell parameter [21] $B_{\max} = 2\sqrt{1 + \widetilde{C}^2}$ can still exceed the classical value 2. This means that the Bell inequality can in principle be violated for arbitrary nonzero $\widetilde{C}$. The optimal violation angles are given by [21]

$$\tan\theta_R = -\widetilde{C}\cot\theta_S, \quad \tan\theta'_R = \widetilde{C}\tan\theta_S,$$

$$\tan\theta_A = \text{sgn}(\cos\theta_R)\sqrt{\frac{\tan^2\theta_S + \widetilde{C}^2}{\widetilde{C}^2\tan^2\theta_S + 1}}, \quad (11)$$

where $\theta_{S,A} \equiv (\theta_L + \theta'_L)/2$. Although it is possible to observe a violation of the Bell inequality under a finite dephasing, the range of angles that can achieve a violation shrinks as $\widetilde{C} \to 0$.

In conclusion, we have investigated the production and detection of electron-hole entanglement based on a QSHI setup. Entangled electron-hole pairs can be created via inter-edge tunneling between the two edges of a single QSHI strip and does not require to fine tune tunneling parameters. The entanglement of such pairs can be inferred from a violation of a Bell inequality based on nonlocal current-current correlations in the four outgoing directions along the edges of the QSHI.

The authors acknowledge stimulating discussions with Daniel Loss. This work was partially supported by the NSF under Grant No. DMR-0840965, Grant No. 228481 from the Simons Foundation, and FAME (an SRC STARnet center sponsored by MARCO and DARPA).

---

Koji Sato, Mircea Trif, and Yaroslav Tserkovnyak

*Department of Physics and Astronomy, University of California, Los Angeles, California 90095, USA*

In this Supplementary Material we give details on the calculation of several quantities presented in the Main Text (MT), such as the derivation of the scattering matrix in Eq. (2) and the evaluation of electron-hole entanglement via the concurrence in Eqs. (5)-(7).

## SCATTERING MATRIX FOR HELICAL STATES

In this section we derive explicitly the general time-reversal invariant scattering matrix in Eq. (2). In Fig. 1 we show a sketch of the scattering process: the incoming states, of amplitude $a_{\alpha,\sigma}$ scatter into the outgoing states of amplitude $b_{\beta,\sigma'}$ with $\alpha, \beta = u, l$ and $\sigma, \sigma' = \uparrow, \downarrow$ in the presence of a constriction (showed in gray). The states at either right or left of the scatterer are related by time-reversal symmetry. In the left (right) part of Fig. 1 we show, in red, the spin-up state helical state, $\psi$ ($\phi$), while in blue we show its time-reversed partner, or spin-down, $T\psi$ ($T\phi$), with $T = i\sigma_y K$ being the time-reversal operator ($\sigma_y$ is the Pauli matrix that act in the spin space and $K$ is the complex conjugation). Thus, the incoming and outgoing states at either the left or the right of the scatterer (red) have their time-reversal partner (blue). The outgoing and incoming states are related via a scattering matrix $S$, i.e. $(\boldsymbol{b}_L, \boldsymbol{b}_R) = S(\boldsymbol{a}_L, \boldsymbol{a}_R)$, with $\boldsymbol{a}_L = (a_{u,\uparrow}, a_{l,\downarrow})$, $\boldsymbol{a}_R = (a_{l,\uparrow}, a_{u,\downarrow})$, and $\boldsymbol{b}_L = (b_{l,\uparrow}, b_{u,\downarrow})$, and $\boldsymbol{b}_R = (b_{u,\uparrow}, b_{l,\downarrow})$. We write this scattering matrix as follows

$$S = \begin{pmatrix} \hat{r} & \hat{t}' \\ \hat{t} & \hat{r}' \end{pmatrix}, \tag{1}$$

with $\hat{r}$, $\hat{t}$, $\hat{r}'$, and $\hat{t}'$ being all $2 \times 2$ matrices and act in the spin space only. The unitarity of $S$ corresponds to the conditions $S^\dagger S = S S^\dagger = 1$ so that $\hat{t}\hat{t}^\dagger + \hat{r}\hat{r}^\dagger = 1$ ($\hat{t}'\hat{t}'^\dagger + \hat{r}'\hat{r}'^\dagger = 1$). Moreover, the time reversal invariance of the scattering matrix $S$ pertains to the condition

$$S^T = \sigma_y S \sigma_y, \tag{2}$$

which results in the following set of equations for the scattering matrix components: $\hat{r}^T = \sigma_y \hat{r} \sigma_y$, $\hat{t}^T = \sigma_y \hat{t}' \sigma_y$, and $\hat{r}'^T = \sigma_y \hat{r}' \sigma_y$. As a consequence, we get $\hat{r} = rI$, $\hat{r}' = r'I$, and

$$\hat{t} = \begin{pmatrix} t_1 & t_2 \\ t_3 & t_4 \end{pmatrix}, \quad \hat{t}' = \begin{pmatrix} t_4 & -t_2 \\ -t_3 & t_1 \end{pmatrix}, \tag{3}$$

with $r$, $r'$, $t_i$ ($i = 1 \ldots 4$) being complex numbers, and $I$ is the $2 \times 2$ identity matrix. From the unitarity, we also obtain:

$$|r|^2 + |t_1|^2 + |t_2|^2 = 1, \tag{4}$$

$$|r|^2 + |t_3|^2 + |t_4|^2 = 1, \tag{5}$$

$$|r'|^2 + |t_1|^2 + |t_3|^2 = 1, \tag{6}$$

$$|r'|^2 + |t_2|^2 + |t_4|^2 = 1, \tag{7}$$

$$t_1 t_3^* + t_2 t_4^* = 0, \tag{8}$$

$$t_1 t_2^* + t_2 t_4^* = 0, \tag{9}$$

$$r t_1^* + r'^* t_4 = 0, \tag{10}$$

$$r^* t_4 + r' t_1^* = 0. \tag{11}$$

From these expressions, we obtain $|t_1| = |t_4|$, $|t_2| = |t_3|$, and $|r'| = |r|$. Writing complex numbers in the polar form, $a = |a|e^{i\phi_a}$, we get several relations between the phases of the scattering matrix elements: $\phi_{t_1} + \phi_{t_4} = \phi_{t_2} + \phi_{t_3} + \pi$, as well as $\phi_r + \phi_{r'} = \phi_{t_2} + \phi_{t_3}$. Let us define $\phi_1 = (\phi_{t_1} - \phi_{t_4})/2$, $\phi_2 = (\phi_{t_2} - \phi_{t_3})/2$, $\phi_3 = (\phi_r - \phi_{r'})/2$, $\phi_4 = (\phi_{t_2} + \phi_{t_3})/2$, as well as $|t_1| = |t_4| \equiv \tau \cos\theta$ and $|t_2| = |t_3| = \tau \sin\theta$, with $\tau$ being a positive real parameter and $\theta \in [0, 2\pi]$. Note that from the above conditions we get that $|r|^2 + \tau^2 = 1$, implying that $0 < \tau < 1$. With these substitutions, and after some straightforward manipulations, we can write $\hat{t} = \tau e^{i\phi_4} Q$, with

$$Q = \begin{pmatrix} \cos\theta e^{i\phi_1} & \sin\theta e^{i\phi_2} \\ \sin\theta e^{-i\phi_2} & -\cos\theta e^{-i\phi_1} \end{pmatrix}. \tag{12}$$

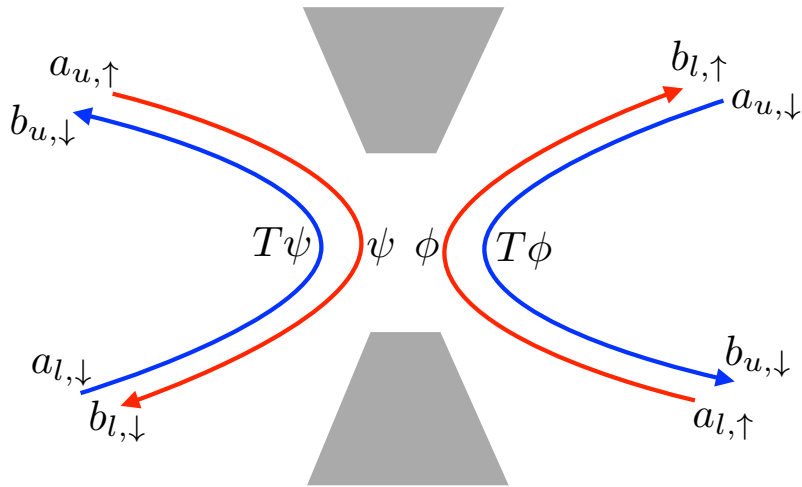

FIG. 1. Sketch of the scattering process presented in terms of the left-right basis. The incoming (outgoing) states are labeled by $a_{\alpha,\sigma}$ ($b_{\alpha,\sigma}$), with $\alpha = l, u$ (lower,upper) and $\sigma = \uparrow, \downarrow$. As stated in the Main Text, we can organize all these states into left (right) states as $\boldsymbol{a}_L = (a_{u,\uparrow}, a_{l,\downarrow})$ [$\boldsymbol{a}_R = (a_{l,\uparrow}, a_{u,\downarrow})$] for the incoming states, and $\boldsymbol{b}_L = (b_{l,\uparrow}, b_{u,\downarrow})$ [$\boldsymbol{b}_R = (b_{u,\uparrow}, b_{l,\downarrow})$] for the outgoing states. The helical states are time-reversal pairs $\psi$ and $T\psi$ ($\phi$ and $T\phi$) on the left (right), with $T = i\sigma_y K$ being the time-reversal operator.

Consequently, we also obtain $\hat{t}' = -\tau e^{i\phi_4} Q^\dagger$. Finally, by using the above findings, we can write for the scattering matrix

$$S = \begin{pmatrix} r e^{i\phi_3} I & -\tau Q^\dagger \\ \tau Q & r e^{-i\phi_3} I \end{pmatrix} e^{i\phi_4}, \tag{13}$$

which is precisely the expression for $S$ depicted in Eq. (2) in the MT (up to the global phase factor $e^{i\phi_4}$). Let us now assume that if we, furthermore, suppose that the system is inversion symmetric in addition to time-reversal invariant, there is an additional constraint on the scattering matrix:

$$S^\dagger = \tau_x S \tau_x, \tag{14}$$

with $\tau_x$ being the Pauli matrix acting in the left-right block (but not on the spin). From this, we get that $e^{2i\phi_4} = 1$ and $\hat{t}^\dagger = \hat{t}$, or $\phi_4 = 0, \pi$ and $\phi_1 = 0$ in the resulting scattering matrix.

Next we analyze the scattering matrix in the regime of weak tunneling between the upper and lower edges (in our picture in Fig. 1 such a processes does not look very natural, but it can be read easily as it covers all possible physical situations). In such a regime, and assuming the scattering matrix is also inversion symmetric, we obtain that $\tau \approx 1$, $r \ll 1$, and $\tau \sin\theta \ll 1$, which pertains to $\theta \approx 0$. In this limit, the Bell inequality acquires the simple expression shown in Eq. (8) in the MT.

## DERIVATION OF THE ELECTRON-HOLE ENTANGLEMENT

In this section we present an explicit derivation of the electron-hole entanglement in our setup in terms of the concurrence. The derivation follows closely the one in Ref. [1], but it is adapted to the helical states (instead of chiral states). We start by writing down the many-body incoming state is written as [1]

$$|\Psi_{\text{in}}\rangle = \prod_\varepsilon a_{u,\uparrow}^\dagger(\varepsilon) a_{l,\downarrow}^\dagger(\varepsilon)|0\rangle = \prod_\varepsilon \frac{i}{2}\sigma_y^{ss'} a_{L,s}^\dagger a_{L,s'}^\dagger|0\rangle, \tag{15}$$

where the energy is restricted to the range $E_F < \varepsilon < E_F + eV$. Here, $|0\rangle$ is the ground state, which is filled up to the Fermi energy for both left- and right-moving incoming states. Although the incoming electrons have to go through the beam splitter on the left, the state $|\Psi_{\text{in}}\rangle$ remains invariant under the unitary splitter in the absence of backscattering. The outgoing many-body state, $|\Psi_{\text{out}}\rangle$, can be found by relating the outgoing states to the incoming states according to Eq. (1) [1]:

$$|\Psi_{\text{out}}\rangle = \prod_\varepsilon i \left[ \left( \hat{r}\sigma_y \hat{r}^T \right)_{12} b_{l,\uparrow}^\dagger b_{u,\downarrow}^\dagger + \left( \hat{t}\sigma_y \hat{t}^T \right)_{12} b_{u,\uparrow}^\dagger b_{l,\downarrow}^\dagger + \left( \hat{r}\sigma_y \hat{t}^T \right)_{ss'} b_{L,s}^\dagger b_{R,s'}^\dagger \right] |0\rangle, \tag{16}$$

up to the overall phase factor $e^{2i\phi_4}$. We make further progress by assuming the weak tunneling regime, $\tau \approx 1$, $|r| \ll 1$ and $\theta \approx 0$, corresponding to the schematics of Fig. 1 in the MT. In this limit we obtain:

$$|\Psi_{\text{out}}\rangle \approx \prod_\varepsilon \left( \sqrt{1-w} - \gamma_{ss'} b_{R,s'}^\dagger b_{L,s}^\dagger \right) |\bar{0}\rangle, \tag{17}$$

where $\gamma \equiv \sigma_y \hat{r}\sigma_y \hat{t}^T = r\tau Q^T$, and $w = \text{Tr}\gamma\gamma^\dagger$ and

$$|\bar{0}\rangle = \prod_\varepsilon b_{u,\uparrow}^\dagger b_{l,\downarrow}^\dagger |0\rangle \tag{18}$$

is the redefined "vacuum state" that includes the completely filled right-moving outgoing states in the energy range $E_F < \varepsilon < E_F + eV$ [1]. Thus, the outgoing many-body state is a superposition of the vacuum state $|\bar{0}\rangle$ and electron-hole excitations (with respect to $|\bar{0}\rangle$). Physically, the scattering results in an orbitally entangled electron-hole pair with electron and hole moving away from each other toward the opposite reservoirs, as depicted in Fig. 1 in the MT. We note also that the weak and strong tunneling tunneling regimes lead to equivalent results, as shown in Ref. [1].

The entanglement of the electron-hole excitations can be quantified by the entanglement of formation $\mathcal{E}$ for a bipartite system [2], which is given by

$$\mathcal{E} = -\text{Tr}_A (\rho_A \log_2 \rho_A). \tag{19}$$

Here, given a density matrix $\rho = |\psi\rangle\langle\psi|$ ($\rho_A = \text{Tr}_B \rho$) for a two-state wave function $|\psi\rangle = \sum_{i,j=1,2} \gamma_{ij} |i\rangle_A |j\rangle_B$, partial traces $\text{Tr}_{A(B)}$ over the first (second) particle are taken. Then, the entanglement of formation reads

$$\mathcal{E} = \mathcal{F}\left( \frac{1}{2} + \frac{1}{2}\sqrt{1 - 4\text{Det}\gamma\gamma^\dagger/(\text{Tr}\gamma\gamma^\dagger)^2} \right), \tag{20}$$

where $\mathcal{F}(x) \equiv -x\log_2 x - (1-x)\log_2(1-x)$. $\mathcal{E}$ is a monotonically increasing function of

$$C \equiv \frac{2\sqrt{\text{Det}\gamma\gamma^\dagger}}{\text{Tr}\gamma\gamma^\dagger}, \tag{21}$$

known as the concurrence. Note that for a purely bipartite system, $\text{Tr}\gamma\gamma^\dagger = 1$, but here we consider a more general case where the entangled part $|\psi\rangle$ of the full wave function is not normalized. $C$ is a bounded function ($0 \le C \le 1$), where $C = 0$ indicates no entanglement, and $C = 1$ corresponds to the maximal entanglement. In general, the entanglement of the state in Eq. (17) is signaled by a finite $C$. In the MT we are quantifying this parameter in terms of the current-current correlations $S_{ij}$ [see Eqs. (5)-(7)], which are measured in an actual experiment.

---